\documentclass[fleqn,usenatbib]{mnras}

\usepackage[T1]{fontenc}

\DeclareRobustCommand{\VAN}[3]{#2}
\let\VANthebibliography\thebibliography
\def\thebibliography{\DeclareRobustCommand{\VAN}[3]{##3}\VANthebibliography}

\usepackage{graphicx}	% Including figure files
\usepackage{amsmath}	% Advanced maths commands
\usepackage{threeparttable}
\usepackage{amsmath}    % Advanced maths commands
\usepackage{amssymb}    % Extra maths symbols
\usepackage[left = `, right = ', leftsub = ``, rightsub = '']{dirtytalk}
\usepackage{multirow}
\usepackage{gensymb}
\usepackage{color}
\usepackage[singlelinecheck=off]{subcaption}

\usepackage{newtxtext,newtxmath}

\usepackage{anyfontsize}

\title[The Cow-like AT\,2024wpp]{Optical evolution of AT\,2024wpp: the high-velocity outflows in Cow-like transients are consistent with high spherical symmetry}

\author[M. Pursiainen et al.]{M. Pursiainen,$^{1}$$^\ddagger$\thanks{E-mail: Miika.Pursiainen@warwick.ac.uk $^\ddagger$ Joint first authorship}
T.~L. Killestein,$^{2}$$^\ddagger$
H. Kuncarayakti$,^{2}$
P. Charalampopoulos,$^{2}$
B. Warwick,$^{1}$
J. Lyman,$^{1}$
\newauthor
R. Kotak,$^{2}$
G. Leloudas,$^{3}$
D. Coppejans,$^{1}$
T. Kravtsov,$^{2}$
K. Maeda,$^{4}$
T. Nagao,$^{2,5,6}$
K. Taguchi,$^{4}$
\newauthor
K. Ackley,$^{1}$
V.~S.~Dhillon,$^{7,8}$
D.~K. Galloway,$^{9}$
A.~Kumar,$^{10,1}$ 
D.~O'Neill,$^{1}$ 
G.~Ramsay,$^{11}$
D.~Steeghs$^{1}$ 
\\
$^{1}$Department of Physics, University of Warwick, Gibbet Hill Road, Coventry CV4 7AL, UK.\\
$^{2}$Department of Physics and Astronomy, University of Turku, Vesilinnantie 5, Turku FI-20014, Finland.\\
$^{3}$DTU Space, National Space Institute, Technical University of Denmark, Elektrovej 327, 2800 Kgs. Lyngby, Denmark.\\
$^{4}$Department of Astronomy, Kyoto University, Kitashirakawa-Oiwake-cho, Sakyo-ku, Kyoto, 606-6 8502, Japan.\\
$^{5}$Aalto University Mets\"ahovi Radio Observatory, Mets\"ahovintie 114, FI-02540 Kylm\"al\"a, Finland.\\
$^{6}$Aalto University Department of Electronics and Nanoengineering, P.O. BOX 15500, FI-00076 Aalto, Finland.\\
$^{7}$ Astrophysics Research Cluster, School of Mathematical and Physical Sciences, University of Sheffield, Sheffield, S3 7RH, UK.\\
$^{8}$ Instituto de Astrof\'{i}sica de Canarias, E-38205 La Laguna, Tenerife, Spain.\\
$^{9}$ School of Physics \& Astronomy, Monash University, Clayton VIC 3800, Australia.\\
$^{10}$ Department of Physics, Royal Holloway, University of London
Egham Hill, Surrey, TW20 0EX, UK\\
$^{11}$ Armagh Observatory \& Planetarium, College Hill, Armagh, BT61 9DG.
}

\date{Accepted XXX. Received YYY; in original form ZZZ}

\pubyear{\the\year{}}

\begin{document}
\label{firstpage}
\pagerange{\pageref{firstpage}--\pageref{lastpage}}
\maketitle

\begin{abstract}
We present the analysis of optical/near-infrared (NIR) data and host galaxy properties of a bright, extremely-rapidly evolving transient, AT\,2024wpp, which resembles the enigmatic AT\,2018cow. AT\,2024wpp rose to a peak brightness of $c=-21.9$\,mag in $4.3$\,d and remained above the half-maximum brightness for only $6.7$\,d. The blackbody fits to the photometry show that the event remained persistently hot ($T\gtrsim20000$\,K) with a rapidly receding photosphere ($v\sim11500$\,km/s), similarly to AT\,2018cow albeit with a several times larger photosphere. $JH$ photometry reveal a NIR excess over the thermal emission at $\sim+20$\,d, indicating a presence of an additional component. The spectra are consistent with blackbody emission throughout our spectral sequence ending at $+21.9$\,d, showing a tentative, very broad emission feature at $\sim5500$\,Å -- implying that the optical photosphere is likely within a near-relativistic outflow. Furthermore, reports of strong X-ray and radio emission cement the nature of AT\,2024wpp as a likely Cow-like transient. AT\,2024wpp is the second event of the class with optical polarimetry. Our $BVRI$ observations obtained from $+6.1$ to $+14.4$\,d show a low polarisation of $P\lesssim0.5$\% across all bands, similar to AT\,2018cow that was consistent with $P\sim0$\% during the same outflow-driven phase. In the absence of evidence for a preferential viewing angle, it is unlikely that both events would have shown low polarisation in the case that their photospheres were aspherical. As such, we conclude that the near-relativistic outflows launched in these events are likely highly spherical, but polarimetric observations of further events are crucial to constrain their ejecta geometry and stratification in detail.
\end{abstract}


%%%%%%%%%%%%%%%%%%%%%%%%%%%%%%%%%%%%%%%%%%%%%%%%%%

%%%%%%%%%%%%%%%%% BODY OF PAPER %%%%%%%%%%%%%%%%%%

\section{Introduction}

\begin{figure*}
    \centering
    \includegraphics[width=0.49\linewidth]{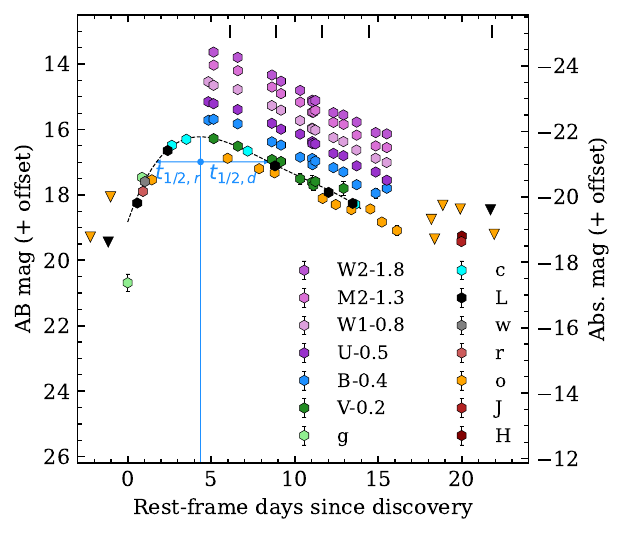}
    \includegraphics[width=0.44\linewidth]{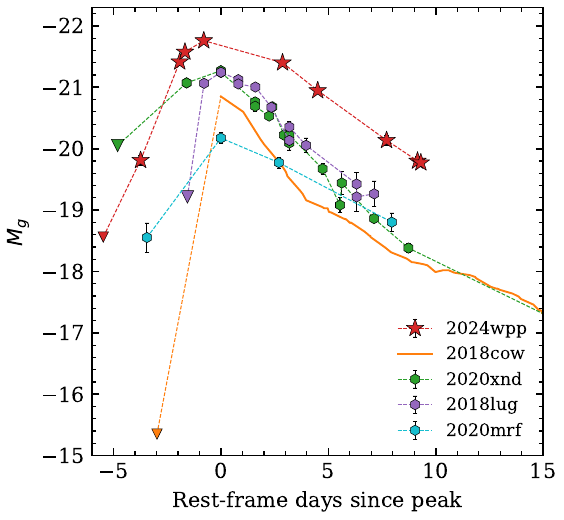}
    \caption{\textbf{Left panel:} Collated photometry of AT\,2024wpp, corrected for Galactic reddening and shown in the transient rest-frame with respect to the ZTF first detection. Overplotted is a polynomial interpolation of ATLAS $c$ and GOTO $L$ bands to determine an approximate peak epoch and timescale -- with these key parameters plotted. The vertical bars refer to the epochs of our spectra, but polarimetry was also obtained for the first four epochs.  Offsets are applied to the \textit{Swift} photometry for clarity. \textbf{Right panel:} comparison of AT\,2024wpp to a number of literature Cow-like transients, with $g$-band photometric coverage: AT\,2018cow~\citep{Perley2018}, AT\,2020xnd~\citep{Perley2021}, AT\,2018lug~\citep{Ho2020}, and AT\,2020mrf~\citep{Yao2022}. Magnitudes are corrected for Galactic extinction, and distance moduli are computed from the host redshift using the same cosmology as for AT\,2024wpp.}
    \label{fig:lightcurve}
\end{figure*}

Modern wide-field, high-cadence transient surveys have revolutionised the field of time-domain astronomy, by allowing the discovery of extragalactic phenomena that evolve faster than expected of standard supernovae (SNe). These were recognised as an abundant population in archival searches of surveys such as the Pan-STARRS1 Medium Deep Survey \citep{Drout2014}, and the Dark Energy Survey \citep{Pursiainen2018, Wiseman2020a}, resulting in large samples of spectroscopically unclassified fast transients. Their light curves evolve too rapidly to be explained purely by the decay of radioactive $^{56}$Ni, the canonical power source of standard SNe, and alternative mechanisms have been sought to explain observed properties. It is now known that many of them are likely Type Ibn/Icn SNe~\citep[e.g.][]{Ho2023}, powered by the interaction between their ejecta and surrounding H-poor circumstellar material (CSM). However, since 2018, a particularly luminous, engine-driven subclass of these events, dubbed \say{Cow-like} transients after their prototype AT\,2018cow, has held the spotlight.

AT\,2018cow is one of the most remarkable extragalactic transients discovered in recent years. It rose to a peak brightness of $\sim-20$.5\,mag in the optical in a mere $2.5$\,d, before hastily declining at $0.2$\,mag/d \citep[e.g.][]{Prentice2018, Perley2018}. For the first $\sim15$\,d, its spectra were consistent with blackbody emission ($\gtrsim20000$\,K)  with some extremely broad emission lines that were indicative of a high expansion velocity of $\sim0.1c$. The subsequent spectra showed the emergence of skewed H and He emission at $v\sim4000$\,km/s, implying the presence of a relatively slowly expanding aspherical ejecta \citep{Margutti2018}. Despite first being discovered in the optical, AT\,2018cow was remarkably luminous across all wavelength regimes from radio to X-rays \citep[e.g.][]{RiveraSandoval2018, Ho2019a, Margutti2018, Kuin2018, Nayana2021}, further supporting the extreme outflows and suggesting a presence of a central source for the X-rays emission.

Following \citet{Margutti2018}, who performed a multi-wavelength analysis of AT\,2018cow, the observed properties can be understood via a central engine  surrounded by a complex aspherical ejecta structure: with near-relativistic polar outflows, and a lower velocity ($\sim3000$\,km/s) equatorial ring/torus. Compact object is strongly favoured for the central engine due to quasi-periodic oscillations (QPOs) in the X-ray \citep{Pasham2021a}, as well as persistent late-time optical to X-ray emission \citep{Sun2022a, Sun2023, Chen2023, Chen2023a, Inkenhaag2023, Migliori2024}. During the first $\sim15$\,d, when the spectra are blue and featureless, the optical emission arises from reprocessed X-ray emission through a photosphere formed in the high-velocity outflow \citep[see also e.g.][]{Piro2020, Uno2020, Calderon2021, Chen2022b}. This photosphere eventually recedes to the equatorial torus closer to the central engine, as the reprocessing becomes less efficient in time and the polar outflow dissipates, thus explaining the emerging narrow emission lines. While the ejecta profile is fairly well understood, the nature of the phenomena themselves is not, and the possible scenarios range from tidal disruption events (TDEs) with intermediate-mass black holes \citep[e.g.][]{Perley2018}, engine-powered SNe \citep[e.g. ][]{Prentice2018} and failed SNe with prompt accretion disk formation \citep{Margutti2018}, to a merger of a black hole and Wolf-Rayet star \citep{Metzger2022}. 
There are now six transients reported in the literature that exhibit similar multi-wavelength properties to AT\,2018cow. These are: AT\,2018lug \citep[ZTF18abvkwla;][]{Ho2020}, CSS161010 \citep{Coppejans2020,Gutierrez2024}, AT\,2020xnd \citep[ZTF20acigmel;][]{Perley2021, Bright2022}, AT\,2020mrf \citep[][]{Yao2022}, AT\,2022tsd \citep{Ho2023a, Matthews2023}, and AT\,2023fhn \citep[][]{Chrimes2023, Chrimes2024}.
Here, we refer to the events as \say{Cow-like} transients after the prototype, AT\,2018cow, but note that these events are also often referred to as luminous fast blue optical transients \citep[LFBOTs;][]{Metzger2022}.

Optical polarimetry is a powerful observational technique that can be used to constrain the geometry of events with electron scattering photospheres, such as SNe \citep[e.g.][]{Wang2008} and TDEs \citep[e.g.][]{Leloudas2022}. Out of the Cow-like events, only the prototype AT\,2018cow has any optical polarimetry, though this shows intriguing results. \citet{Maund2023} report high ($\lesssim7$\%) wavelength-dependent polarisation at $+5.7$\,d post-explosion that declined rapidly, attributed to a shock breakout through an optically thick, physically thin disk. The high polarisation is supported by multi-epoch spectral polarimetry obtained at the 2.3m Bok telescope, between phases $+6.0-8.9$\,d \citep{Smith2018a}. The last two epochs of Bok polarimetry ($\gtrsim8$\,d) also constrained the polarisation to be consistent with Milky Way (MW) interstellar polarisation (ISP) with an error of $\delta P\sim0.1$\%, implying that AT\,2018cow showed zero polarisation and is consistent with a high degree of spherical symmetry at this phase. Obtaining polarimetry of other Cow-like transients is crucial to investigate the geometry of the two ejecta components, but as these events have an intrinsically low rate \citep[$\lesssim400$\,Gpc$^{-3}$yr$^{-1}$, $\lesssim0.6$\% of core-collapse SNe;][]{Coppejans2020}, they are typically distant and too faint for high-quality polarimetry to be obtained.

AT\,2024wpp, the newest candidate for the class, was first identified in ZTF survey data on 2024 Sept 26~\citep{Ho2024}, showing a fast rise of $\sim3$\,mag in a day. Subsequent follow-up observations showed strong X-ray emission in \textit{Swift}~\citep{Srinivasaragavan2024} and \textit{NuSTAR}~\citep{Margutti2024}, strong radio emission at 10 and 15\,GHz \citep{Schroeder2024} and verified the transient's association with a diffuse, extended host galaxy at a separation of $3.1\arcsec$ at redshift $z=0.0868$~\citep[][approx 5.2\,kpc]{Perley2024}. With a peak brightness of nearly $-22$\,mag, rapid evolution timescale, and strong X-ray and radio detections, AT\,2024wpp likely belongs to the herd of Cow-like transients.

In this paper, we present the analysis of optical photometry, spectroscopy and polarimetry, as well as host galaxy properties of AT\,2024wpp. The paper is structured as follows: In Section \ref{sec:observations}, we introduce the datasets and how they were reduced, in Sections \ref{sec:LC} -- \ref{sec:impol} we present the analysis of the observations and discuss the implications for the class of Cow-like events, and in Section \ref{sec:conclusions} we conclude our findings. Throughout this paper, we assume the \citet{PlanckCollaboration2020} $\Lambda$CDM cosmology. Using the redshift $z=0.0868$~\citep{Perley2024}, this corresponds to a distance modulus to the host of AT\,2024wpp of $\mu=38.06$\,mag. All magnitudes are given in the AB system. The phases in this paper are given in the transient rest frame relative to the ZTF first detection at $\mathrm{MJD}=60578.44$ \citep{Ho2024}. 

\section{Observations and Data Reduction}

\begin{figure}
    \centering
    \includegraphics[width=0.98\linewidth]{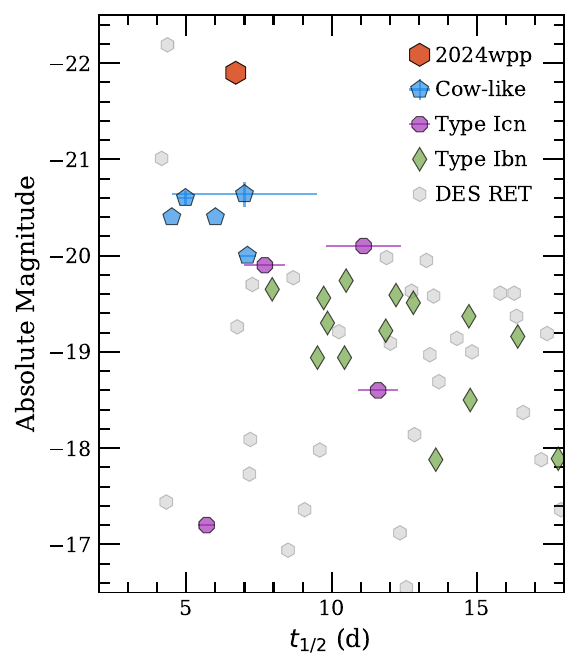}
    \caption{Absolute magnitude and time above half-maximum brightness for selected populations of fast-evolving extragalactic transients. The values for AT\,2024wpp are estimated from the light curve fit presented in Figure \ref{fig:lightcurve}. The comparison objects are shown in either observer-frame $g$- or $r$-band, as provided in the literature. The values are collected from following source: Cow-likes \citep{Perley2018,Ho2020, Perley2021, Yao2022, Ho2023a}; Type Icn SNe \citep{Pellegrino2022}, Type Ibn SNe \citep[ZTF Bright Transient Survey;][]{Fremling2020, Perley2020} and the Dark Energy Survey sample of rapidly evolving transients \citep[RETs;][]{Pursiainen2018}. Uncertainties are shown if available. The bands of individual RETs ($g$, $r$, $i$ or $z$) were chosen to be the closest to the GOTO $L$/ATLAS $c$ used for AT\,2024wpp in their respective rest frames.}
    \label{fig:M_t}
\end{figure}

\label{sec:observations}
We collated all available photometry on AT\,2024wpp.
ATLAS~\citep{Tonry2018,Smith2020} $o$- and $c$-band photometry was retrieved via the ATLAS Forced Photometry Server~\citep{Shingles2021}, and combined per-quad using the inverse-variance-weighted mean of forced flux, with sigma clipping to remove discrepant points. Upper limits were also re-computed based on the stacked fluxes and their uncertainties, with 5$\sigma$ limits being used in practice. Photometry from the Gravitational-wave Optical Transient Observer (GOTO; \citealt{Steeghs2022, Dyer2024}) in the $L$-band was retrieved via the GOTO Forced Photometry service~(Jarvis et al., in prep.). We also collated all publicly-available photometry from relevant Transient Name Server\footnote{\url{https://www.wis-tns.org/}} (TNS) AstroNotes and discovery reports. 
The Neil Gehrels \textit{Swift} Observatory (\textit{Swift}) observed AT\,2024wpp as part of a number of target-of-opportunity programs (PIs: Margutti, Srinivasaragavan, Brown, Coughlin). The images from \textit{Swift} Ultraviolet/Optical Telescope \citep[UVOT;][]{Roming2005} were reduced, and photometry in $W1,M2,W2,U,B,V$ bands was performed following the methodology of \citet{Charalampopoulos2024}. We do not consider \textit{Swift} XRT observations in this work. Finally, we obtained one epoch of near-infrared (NIR) $JH$-band imaging with the Nordic Optical Telescope near-infrared Camera and spectrograph (NOTCAM) instrument mounted on the Nordic Optical Telescope (NOT) at $+20$\,d. Observations were taken with a 9-position box dither pattern, reduced with standard procedures, and measured with aperture photometry, which was then calibrated using 2MASS stars in the field.
Photometry is corrected for Galactic extinction assuming $E(B-V)= 0.0253\pm0.0025$\,mag \citep{Schlafly2011} and $R_V=3.1$, using the filter curves hosted at the SVO Filter Profile Service~\citep{Rodrigo2012,Rodrigo2020}. In lack of evidence for notable host extinction, we do not account for it in our analysis.

We obtained a sequence of optical spectra of AT\,2024wpp using the Kyoto Okayama Optical Low-dispersion Spectrograph with optical-fiber Integral Field Unit~\citep[KOOLS-IFU;][]{Matsubayashi2019} instrument on Seimei Telescope and  Alhambra Faint Object Spectrograph and Camera (ALFOSC), mounted on the Nordic Optical Telescope (NOT). The observation logs, with salient details about the data, are provided in Table \ref{tab:spectral_log}.
The spectrum at $+5.8$\,d taken with KOOLS-IFU was reduced using the Hydra package in IRAF~\citep{Barden1994} and a reduction software developed for KOOLS-IFU data\footnote{\url{http://www.o.kwasan.kyoto-u.ac.jp/inst/p-kools/}}, following the standard procedures (bias, sky subtraction, wavelength calibration with Arc lamps, and relative flux calibration). We used VPH-blue grism with a spectral resolution of $\sim 500$; while it officially gives the wavelength coverage of 4100–8900\,Å, we cut the low-SNR portions of the spectrum, resulting in 4600--8000\,Å in the final spectrum.
Optical spectra obtained with NOT/ALFOSC starting at $+6.1$\,d post-detection were reduced with PyNOT\footnote{\url{https://github.com/jkrogager/PyNOT}} reduction pipeline. Spectra are corrected for Galactic reddening using the \citet{Fitzpatrick1999} dust law, assuming $E(B-V)= 0.0253\pm0.0025$\,mag \citep{Schlafly2011} and $R_V=3.1$. We do not perform absolute flux calibration on the spectra with contemporaneous photometry owing to sparse multi-colour photometry available, but deem the relative flux calibration to be more than adequate to discuss the overall spectroscopic evolution. 

We also obtained a sequence of multi-epoch $BVRI$ polarimetry with NOT/ALFOSC. The imaging polarimetry was reduced with a custom-built reduction pipeline based on \texttt{photutils}~\citep{Bradley2024}, using the reduction steps detailed in \citet{Pursiainen2023a}. The observations were taken during waxing moon, between lunar illumination 0\,--\,53\%, and thus lunar phase should not influence the results~\citep{Pursiainen2023a}. The details of our polarimetry are given in Table \ref{tab:impol_log}.

\begin{figure}
    \centering
    \includegraphics[width=0.49\textwidth]{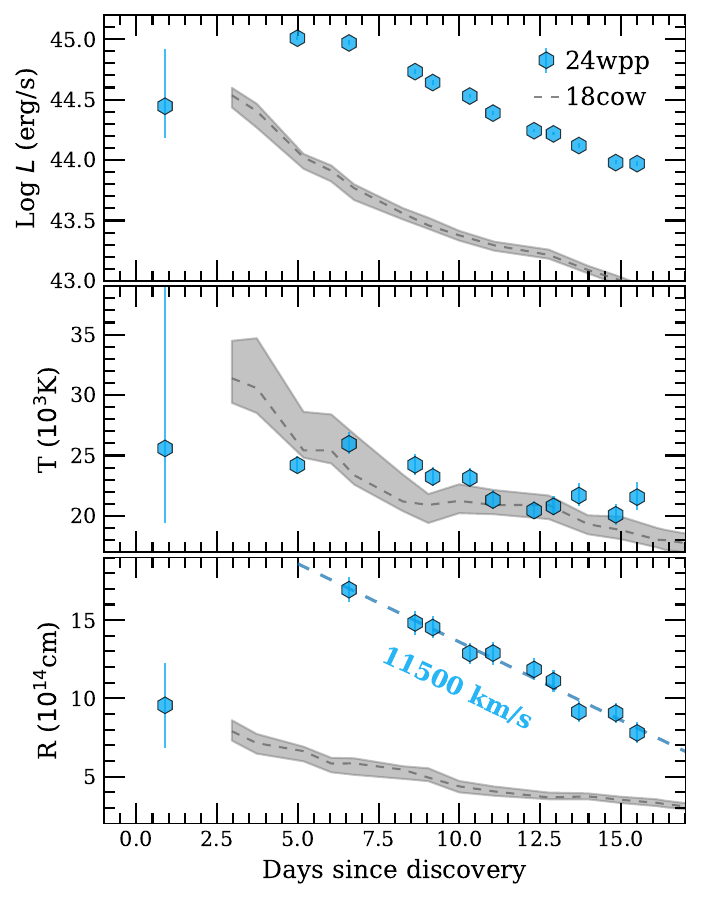}
    \caption{The evolution of bolometric luminosity, temperature and radius of AT\,2024wpp based on the blackbody fits presented in Figure \ref{fig:BB_fits} against AT\,2018cow \citep{Perley2018}. The temperature follows AT\,2018cow closely and shows a persistently high value. The evolution of radius is similar to AT\,2018cow in that it recedes rapidly, but the photosphere of AT\,2024wpp is significantly larger, resulting in higher peak luminosity. The optical photosphere of the event appears to recede at $v\sim11500$\,km/s, as shown by the dashed line.}
    \label{fig:TR_evo}
\end{figure}

\section{Light Curve Evolution}
\label{sec:LC}

The collated light curves of AT\,2024wpp are presented in Figure~\ref{fig:lightcurve}. The time of peak light and corresponding peak (absolute) magnitude are estimated via fitting Chebyshev polynomials to GOTO $L$ and ATLAS $c$-band photometry -- jointly providing the best-sampled light curve. Whilst there is a slight mismatch in bandpass between these two filters, considering the extreme blue colour of the source, and the bandpass differing only in the red, this treatment is adequate for the level of analysis we do here. The fit implies a rise to peak in the transient rest frame of $4.3\pm0.1$\,d, with uncertainties estimated via bootstrap resampling. In the absence of any tightly-constraining non-detections, we assume the ZTF discovery is the explosion time. While the rise time is thus formally a lower limit, we note that given the fast rise of 3\,mag in a day post-discovery \citep{Ho2024}, the explosion epoch has to be very close in time. The peak magnitude based on the fit is $16.3$, corresponding to an absolute magnitude of $-21.9$ in ATLAS $c$/GOTO $L$ under the assumed distance modulus, and correcting for Galactic extinction of approximately 0.078 mag. The post-peak decline of AT\,2024wpp is rapid based on the fit evaluated at +10\,d: 0.24 mag\,$\text{d}^{-1}$ in GOTO $L$/ATLAS $c$. Based on the measured peak time and this peak magnitude, we compute the time above half-maximum ($t_{1/2}$) of the light curve using our fitted interpolant by computing the elapsed times before and after peak that the transient flux decreases by a factor of 2. This yields $t_{1/2, r} = 2.6\,\text{d}$ and $t_{1/2, d} = 4.1\,\text{d}$ for the rise and decline timescales respectively -- implying an overall $t_{1/2}$ of 6.7\,d in these bands.

In terms of the light curve evolution, AT\,2024wpp lies firmly among other published Cow-like events, showing a high peak luminosity and rapid light curve evolution \citep[$M\gtrsim-20$, $t_{1/2}\lesssim7$\,d;][]{Ho2023}. We present a comparison of AT\,2024wpp to other Cow-like transients from the literature in the right panel of Figure~\ref{fig:lightcurve}. Although a full light-curve driven comparison is left to future works with more comprehensive photometric datasets, the overall duration of AT\,2024wpp is more comparable to AT\,2020xnd~\citep[][$t_{1/2}=5.6\pm1.6$ d]{Perley2021} than either AT\,2018cow or AT\,2018lug, however markedly more luminous by $\sim0.8$\,mag. The $(J-H)$ colour of $0.18 \pm 0.12$ mag in AT\,2024wpp is relatively bluer compared to that of AT\,2018cow at a similar phase, $(J-H) \sim 0.30 \pm 0.04$ mag \citep[e.g.][]{Perley2018}, although these are consistent with each other within the uncertainties.

In comparison, no other known phenomena exhibit such properties as a population. To emphasise this, in Figure~\ref{fig:M_t} we compare the luminosities and timescales of the populations of rapidly-evolving transients with AT\,2024wpp. Type Ibn and Icn SNe occasionally exhibit luminosities of up to $\sim-20$\,mag \citep[e.g. Icn SNe 2021csp and 2021ckj;][]{Fraser2021, Perley2022, Nagao2023}, but are typically fainter and slower as a population \citep[see e.g.][]{Hosseinzadeh2017, Pursiainen2023b}, and their spectra are dominated with line features arising from H-poor CSM, which AT\,2024wpp does not exhibit. Similarly, the luminous fast-cooling transients, such as \say{Dougie} \citep{Vinko2015}, are extremely luminous ($\sim-22$\,mag) in nature, but they are clearly slower evolving \citep[$t_{1/2}\sim20$\,d;][]{Nicholl2023} than Cow-like events, and occur solely in passive environments, inconsistent with star-forming host of AT\,2024wpp \citep{Perley2024}. 

To investigate the properties of the photosphere, we fit the multi-band photometry of AT\,2024wpp with a simple blackbody function. The resulting temperature, radius and bolometric luminosity curves are presented in Figure \ref{fig:TR_evo} and the fits themselves in Figure \ref{fig:BB_fits}. AT\,2024wpp a similar evolution to AT\,2018cow. The event exhibits a high temperature of $T\gtrsim20000$\,K throughout, accompanied by a rapidly receding photosphere ($v\sim11500$\,km/s) after peak brightness -- evolution associated with AT\,2018cow \citep[e.g. ][]{Prentice2018, Perley2018}. The photosphere of AT\,2024wpp, however, has a several times larger radius than AT\,2018cow, making it also significantly brighter. Furthermore, while the fit to the early ZTF $gr$ epoch at $+0.9$\,d is uncertain, it clearly implies that the event was hot, but small at the time, and that the photosphere was still expanding as the event grew brighter.

We note that the blackbody fits of the \textit{Swift} six-band photometry appear to show a slight excess in the UV over the optical, despite being largely consistent with thermal emission (see Figure \ref{fig:BB_fits}). This could be caused by significant downscattering of X-ray emission, as proposed for AT\,2018cow \citep{Margutti2018}. \citet{Maund2023}  argue that a resulting excess of blue emission could explain the wavelength dependency of the early polarisation of AT\,2018cow. The observed polarisation peaks towards the red -- a natural consequence if the polarisation carried by the thermal emission is diluted in the blue, thanks to the addition of downscattered unpolarised emission. 

Finally, we consider our epoch of NIR photometry given our inferred blackbody parameters and evolution, in light of suggested dust echoes in Cow-like transients~\citep[e.g.][]{Metzger2023}. Most notably, AT\,2018cow showed a NIR excess at similar epoch \citep{Perley2018}, though only $H$-band photometry had continuous coverage, thus little information on the temperature evolution of this component is available. Assuming a constant temperature ($\sim 21,000$K) beyond +15\,d, and a linearly receding radius, the hot blackbody component alone cannot explain the observed NIR photometry. The extrapolated radius at this epoch is $3.6^{+0.9}_{-0.8}\times10^{14}$\,cm, and we estimate AT\,2024wpp has brightness $J=H=20.6$ mag at +20\,d under these assumptions. This is $\sim$1 magnitude fainter than the observed NOTCAM points ($J=19.43$ mag, $H=19.25$ mag, corrected for extinction) implying a NIR excess arising from an additional SED component must be present. The optical spectrum remains blue at +21.8\,d indicating it remains hot, yet the NIR colour is red (brighter in $H$ than $J$), which is inconsistent with a single thermal component, where we would be sampling the Rayleigh-Jeans tail in $JH$. 
This observed excess is best explained by an additional cooler component present, with properties consistent with those predicted by a dust echo.

\begin{figure}
    \centering
    \includegraphics[width=\linewidth]{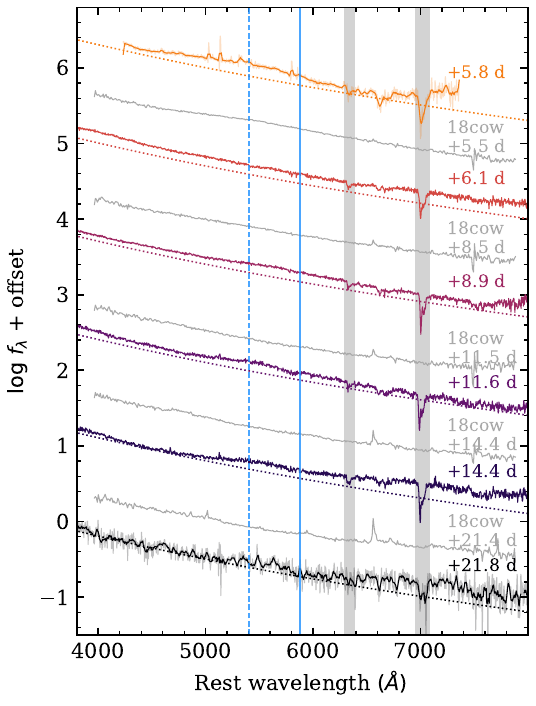}
    \caption{Spectral series of AT\,2024wpp, spanning $+5.8$ to $+21.8$ days post-discovery in the transient rest frame. Spectra are plotted in the transient rest frame, and corrected for Galactic reddening. Shaded bands correspond to the most prominent telluric absorption lines. Interspersed between the AT\,2024wpp spectra are the spectra of AT\,2018cow presented in \citet{Prentice2018}, retrieved via WISeREP~\citep{Yaron2012}. The solid line corresponds to \ion{He}{I} $\lambda$5876 at rest, with the dashed line denoting this ion at a velocity of $\sim0.08c$. We also overplot a 23,000\,K blackbody below each spectrum of AT\,2024wpp to guide the eye. The first and final spectra are smoothed with a 30Å boxcar filter for clarity of visualisation, and cosmic rays are interpolated over.}
    \label{fig:spectral-series}
\end{figure}

\section{Spectral Series and Evolution}
\label{sec:spec}

As shown in Figure~\ref{fig:spectral-series}, the spectra of AT\,2024wpp are consistent with AT\,2018cow during our spectral series -- blue and featureless throughout with no narrow emission or absorption lines present from the transient. The fact that our spectra remain consistently blue and featureless up to the end of the spectral sequence at $\sim+21$\,d is a strong indicator for the presence of a central heating source in AT\,2024wpp as expected for Cow-like transient, as without one the ejecta would cool quickly and ejecta-associated lines would emerge. Further, hints of extremely broad emission lines are visible in our spectra, as has been reported for Cow-like transients previously~\citep[e.g.][]{Perley2018,Margutti2018} -- with their spectra showing velocities $v\sim0.1c$ at early times. The most prominent, a broad spectral feature around 5500\,Å, is seen in AT\,2024wpp growing in strength as the light curve of AT\,2024wpp declines. Given prior Cow-like events have predominantly shown emission lines of H and He, this feature could be associated with \ion{He}{I} $\lambda$5876, highly broadened and offset from rest by $\sim0.08$c. The next-closest possibility is H$\alpha$ which would require a velocity of $\sim0.3c$, and as such we consider it less plausible. While the required helium velocity is high, it is still reasonable given the high-velocity outflows invoked in, for instance, AT\,2018cow and CSS161010, to explain their observed high-energy properties \citep[e.g.][]{Margutti2018,Coppejans2020}. Further, recent work on CSS161010~\citep{Gutierrez2024} identified \ion{He}{II} emission at comparable phase, with velocities of $\sim0.1c$. Although this is different to the \ion{He}{I} identification, it demonstrates He can be found at comparable velocities. This indeed hints that the photosphere at the phases covered by our spectral series lies within the near-relativistic outflow suggested for Cow-like events \citep[e.g.][]{Margutti2018}. Following other Cow-like event, later spectroscopy of the event might show emerging narrow emission lines that are possibly skewed, as in AT\,2018cow \citep{Margutti2018}.

\section{Host Galaxy Properties}
\label{sec:host}

\begin{figure}
    \centering
    \includegraphics[width=\linewidth]{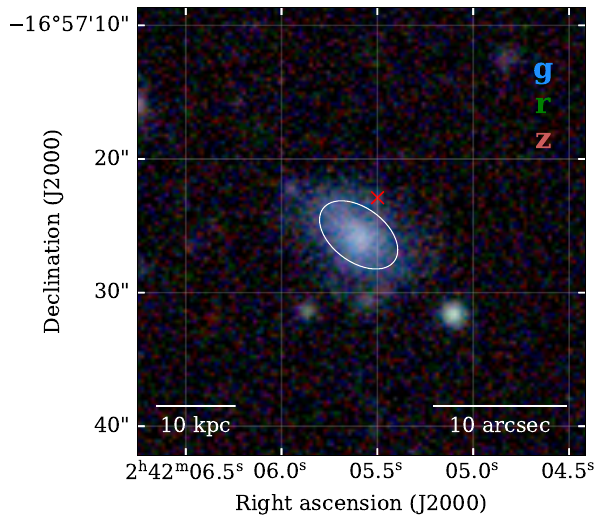}
    \caption{DESI Legacy Surveys $grz$ cutout of the explosion site of AT\,2024wpp, marked with the red cross. The white ellipse marks the contour that encloses 50\% of the host flux. Data are plotted with logarithmic scaling to emphasise the underlying faint host galaxy.}
    \label{fig:host_stamp}
\end{figure}

AT\,2024wpp occurred on the outskirts of a blue, diffuse galaxy at $z=0.0868$ \citep{Perley2024} as shown in Figure \ref{fig:host_stamp}. To quantify the galaxy properties, we used the  $griz$ images as well as the Wide-field Infrared Survey Explorer (WISE) $W1$ and $W2$ fluxes from the 10th DESI Legacy Survey \citep{Dey2019} data release\footnote{\url{www.legacysurvey.org/dr10}}. The multi-band data was fitted with the \texttt{CIGALE} software \citep{Boquien2019} following details presented in \citet{Warwick2024}. \texttt{CIGALE} computes a composite spectral energy distribution (SED) from different galaxy components. The galaxy templates used were created with a delayed star formation history, utilising the simple stellar population models from \citet{Bruzual2003}. 

The host galaxy is found to be relatively faint at $g=-17.91$, $r=-18.38$, $i=-18.61$ and $z=-18.70$, and the \texttt{CIGALE} fit resulted in a stellar mass of $1.15\pm0.55\times10^9M_\odot$. The host of AT\,2024wpp appears to be similar to that of AT\,2018cow, which was found to have a stellar mass of $\sim1.5\times10^{9}$\,M$\odot$ with moderate star-formation rate of $\mathrm{SFR}\sim0.2$\,M$_\odot$/year \citep{Perley2018, Lyman2020}. While we consider the estimated stellar mass of the host to be robust, any SFR estimate would be unreliable in the absence of UV coverage to capture the young, hot stellar populations, and we leave the detailed investigation of the star formation history to future studies with improved datasets. However, we note that the host galaxy of AT\,2024wpp is likely star-forming. The host is visibly blue in Figure \ref{fig:host_stamp}, implying contribution from some young stellar populations. The galaxy would also be remarkably low-mass for an isolated passive galaxy. Further, the presence of narrow emission lines has been reported with line ratios that imply star-formation \citep{Perley2024}. As such, we conclude that the host of AT\,2024wpp is similar to that of AT\,2018cow and comparable to the Large Magellanic Cloud (LMC).

We also investigated if the location of AT\,2024wpp is common in the context of Cow-like transients. For this purpose, we estimated the half-light radius $r_\mathrm{50}$ of the galaxy towards AT\,2024wpp in $g$-band using \texttt{starmorph} software \citep{Rodriguez-Gomez2019} similarly to \citet{Chrimes2023}. The estimated half-light ellipse of the galaxy is shown in Figure \ref{fig:host_stamp}. AT\,2024wpp occurred $5.1$\,kpc from the host nucleus, which is significantly further away than the $r_\mathrm{50}=3.3$\,kpc in the direction of AT\,2024wpp. The Cow-like transient most distant from its host, AT\,2023fhn \citep{Chrimes2023}, was found to be either at $16.5$\,kpc if it is associated with the large near-by spiral galaxy, or at $5.4$\,kpc if it occurred in the smaller satellite. While the distance of AT\,2024wpp is comparable if one assumes the satellite as the host of AT\,2023fhn, in normalised offsets ($r_\mathrm{n}=D/r_\mathrm{50}$), AT\,2024wpp is significantly closer to its host: $r_\mathrm{50}=1.5$ while for AT\,2023fhn the values are $3.7$ and $3.6$ (the spiral and satellite, respectively). The values for AT\,2023fhn are estimated in F555W, which at the redshift ($z=0.24$) is comparable to $g$-band for AT\,2024wpp. As such, we conclude that, while AT\,2024wpp is relatively distant from its host for the class, it is not an outlier like AT\,2023fhn. 

%sfr 0.010 +- 0.025 solar mass a year, stellar mass 1.15(+-0.55)e9 solar masses, metallicity 0.0011+-0.0019.

%\begin{table}
%\def\arraystretch{1.1}%
%\setlength\tabcolsep{8pt}
%\centering
%\fontsize{9}{11}\selectfont
%\caption{Log of NOT/ALFOSC imaging polarimetry observations presented in this paper.}
%\label{tab:host_prop}
%\begin{tabular}{c c}
%\hline
% Band & Mag \\
% $g$   & -17.91 \\
% $r$   & -18.38 \\
% $i$   & -18.61 \\
% $z$   & -18.70 \\
% $W1$  & xx \\
% $W2$  & xx \\
%\hline
%\end{tabular}
%\end{table}

\begin{figure*}
    \centering
    \includegraphics[width=0.8\textwidth]{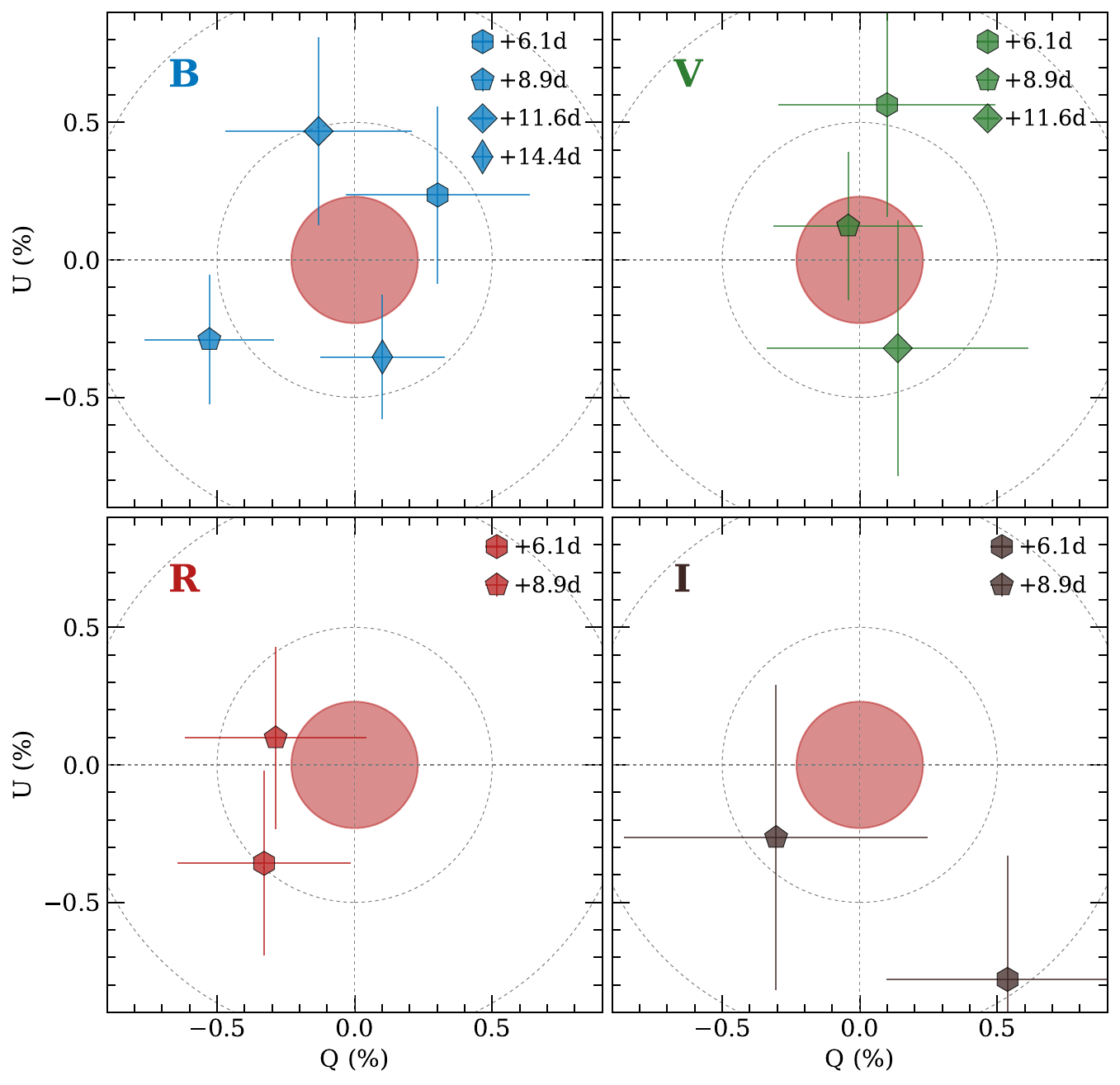}
    \caption{The Stokes $Q$ -- $U$ planes of NOT/ALFOSC $BVRI$ polarimetry of AT\,2024wpp. The event shows low polarisation ($\lesssim0.5$\%) and is consistent with zero polarisation regardless of the phase and band. The possible region of Galactic ISP ($\gtrsim0.23$\%) is highlighted with a red, filled circle. The dashed lines mark $Q,U = 0$\% and $P = 0.5,1.0$\%.  }
    \label{fig:QU}
\end{figure*}

\section{Multi-epoch Imaging Polarimetry}
\label{sec:impol}

Our NOT/ALFOSC imaging polarimetry is presented in Figure \ref{fig:QU} and Table \ref{tab:impol_results}. The polarimetry is low at $P\lesssim0.5$\%  regardless of wavelength and phase of observation, and the values are consistent with zero polarisation. The low value does not seem to be the result of destructive interstellar polarisation (ISP). The field of view of the NOT polarimetry does not include any bright stars, and therefore a direct measurement of MW ISP is not possible, but indirect means can be used to constrain its significance. The line-of-sight colour excess due to Galactic dust is very low, at $E(B-V)= 0.0253\pm0.0025$\,mag \citep{Schlafly2011}. Using the empirical Serkowski relation $\mathrm{P_\mathrm{ISP}}\leq9\times E(B-V)$ \citep{Serkowski1975}, the MW dust content corresponds to an ISP of no greater than $P\sim0.23$\%, well within our observational uncertainties. Low ISP is further supported by stars catalogued in \citet{Heiles2000} near the line-of-sight of the event. There are four stars within $3\deg$, all with $P<0.15$\%. The plausible extent of Galactic ISP based on the Serkowski relation is shown in Figure \ref{fig:QU} with a filled red circle. Similarly, we can constrain the effect of host galaxy ISP using indirect methods. The spectra and photometry appear to be consistent with blue continua, and while the NIR excess might indicate a dust echo, there are no obvious signs of dust scattering in the optical (see Figures \ref{fig:spectral-series} and \ref{fig:BB_fits}). We also see no \ion{Na}{ID} absorption from the host galaxy, often used to estimate extinction in the host \citep{Poznanski2012}, indicating a low dust content. Finally, we note that it would be unlikely to have such a magnitude and orientation of host galaxy ISP that would be completely destructive, and cause the intrinsically polarised light from the event to appear consistent with $P=0$\%. As such, we conclude that the total ISP must be low, and that the event shows intrinsically low polarisation. In the case of an oblate photosphere, $P\sim0.5$\% implies an ellipticity of $b/a\sim0.9$ \citep{Hoflich1991}, and the polarimetry is consistent with a high degree of spherical symmetry.

AT\,2024wpp is only the second Cow-like transient with optical polarimetry. The prototypical AT\,2018cow showed high, wavelength-dependent polarisation of up to $\sim7$\% at $+5.7$\,d post-explosion, attributed to a shock breakout in a disk of material \citep{Maund2023}. Our observations of AT\,2024wpp do not show such a high value, but since our polarimetric sequence started $+6.1$\,d after the first reported detection in ZTF \citep{Ho2024}, it is possible that this phase was not covered by our observing campaign especially given the events show a range of evolution timescales despite being always fast (see Figure \ref{fig:lightcurve}). However, from $\sim+8$\,d onwards AT\,2018cow is perfectly consistent with the inferred MW ISP to high accuracy \citep[$\delta P\sim0.1$;][]{Smith2018a}, implying zero polarisation, similar to AT\,2024wpp at a comparable epoch. \citet{Maund2023} also report a consequent increase of polarisation up to $P\sim1.5$\% at $\sim+13$\,d in the bluer wavelengths, which we rule out for AT\,2024wpp. 

In the context of SNe, a low polarisation normally implies a near-spherical photosphere \citep[e.g.][]{Wang2008}. However, it is typically not possible to confirm it with only one event, as the projection of an aspherical geometry can be circular on the sky and thus result in low polarisation.  Now with AT\,2024wpp, both events with optical polarimetry show low polarisation at a similar phase and their photospheres at this epoch are consistent with high spherical symmetry. At face-value, this is at odds with the conclusions made for other Cow-like events that imply a complex stratified ejecta profile, producing an intrinsically aspherical photosphere. \citet{Margutti2018} used multi-wavelength data to constrain the location of the optical photosphere of AT\,2018cow to be in the near-relativistic polar outflow until about $\sim15$\,d post-discovery. The photosphere then receded to a dense equatorial torus, viewed close to edge-on, with the torus reprocessing hard X-rays arising from a central engine. The early ($+5.7$\,d) high polarisation of AT\,2018cow that was attributed to a disk-like geometry also supports a large viewing angle \citep{Maund2023}. Assuming that AT\,2024wpp is powered by a similar mechanism and has an overall similar ejecta geometry to AT\,2018cow, the optical polarimetry is likely taken when the optical photosphere is located in the high-velocity outflow as supported by the spectral properties (see Section \ref{sec:spec}). There are two possible reasons why the two events show low polarisation: 1) There is a preferential viewing angle to discover these phenomena related to the outflow direction so that projection of intrinsically aspherical photospheres would appear circular or 2) their outflows are at least nearly spherical.  

A preferential viewing angle along the outflow axis could cause low polarisation. For instance, the low polarisation of the jetted TDE AT\,2022cmc was argued to be consistent with viewing the relativistic jet directly pole-on \citep{Cikota2023}. However, despite the complex ejecta geometry, there is no discussion in the literature on a preferred viewing angle at which to discover Cow-like events, and instead, they appear to be observed at a range of orientations. The prototype AT\,2018cow was likely viewed close to edge-on based on both spectroscopic \citep[e.g.][]{Margutti2018} and polarimetric properties \citep{Maund2023}, but on the other hand, the minute-timescale optical variability in AT\,2022tsd has been used to suggest it is viewed more face-on  \citep{Ho2023a}.  As such, there is no \textit{a priori} reason that suggests a particularly preferential viewing angle.  In fact, if Cow-like events are discovered at random inclinations, it would be more likely to view them equator-on rather than pole-on. As such, it is unlikely that both AT\,2024wpp and AT\,2018cow would have been viewed from a same specific inclination needed to explain the low observed polarisations if the photospheres were aspherical. Therefore, the fact that both events show low polarisation at $\gtrsim+8$\,d implies that the optical photospheres in the near-relativistic outflows, often referred to as \say{polar}, show a high degree of spherical symmetry. This in turn is a strong indication that the outflows themselves are at least nearly spherical in nature. 

Such a configuration is consistent with the deduced ejecta geometry in AT\,2018cow, as the lower-density ejecta profile in the polar direction only requires a separate dense equatorial disk/torus. While the torus undoubtedly affects the high-velocity outflow, given the velocity difference between the outflow ($\sim0.1c$) and the torus ($\sim4000$\,km/s), it is possible that the outflow could flow around it. The polarimetric evolution of AT\,2018cow can be directly understood under such a scenario. The high initial polarisation could still arise from photosphere formed in a disk/torus of material, but once the photosphere carried by the high-velocity, spherical outflow overtakes it, the polarisation would suddenly decrease. A similar scenario has also been modelled in the context of SN surrounded by a disk of CSM \citep{McDowell2018}, which has already been invoked to explain the spectroscopic and polarimetric evolution of the superluminous SN\,2018bsz \citep{Pursiainen2022}. The only genuine constraint for such a scenario is that in case the disk/torus is geometrically-thick and covers a large opening angle, the outflow could possibly be funnelled towards the poles and be suppressed in the equatorial direction. As such, the equatorial torus must be reasonably constrained in size to allow the outflow to largely engulf it, as proposed. However, polarimetry of a sample of Cow-like events from near explosion to late time is needed to investigate their ejecta stratification and overall geometry in detail.

\section{Conclusions}
\label{sec:conclusions}

Here we have presented the analysis of optical/NIR data as well as host galaxy properties of the bright, rapidly-evolving transient AT\,2024wpp, and multiple angles support it being a Cow-like event.  Its extreme brightness of $-21.9$\,mag, fast-evolving light curve ($t_{1/2}=6.7$\,d), persistently hot ($T\gtrsim20000$\,K), but rapidly receding photosphere ($v\sim11500$\,km/s) and the NIR excess at $\sim+20$\,d are consistent with the population, but are extreme outliers for any other class of extragalactic events. Our NOT spectra are consistent with blackbody emission all the way to $+22$\,d implying the presence of a central heating source, as well as showing tentative extremely broad emission features, consistent with a near-relativistic outflow. The host galaxy is similar to that of AT\,2018cow with a stellar mass of $1.15\pm0.55\times10^9M_\odot$, and while AT\,2024wpp occurred on the outskirts of its host ($5.1$\,kpc), it is still clearly in the galaxy at a normalised offset $r_\mathrm{n}=r/r_\mathrm{50}=1.5$. Furthermore, the event has shown strong X-ray and radio emission \citep{Margutti2024, Schroeder2024,Srinivasaragavan2024} as expected of the class. As such, AT\,2024wpp appears to be part of the growing herd of Cow-like transients.

AT\,2024wpp is only the second Cow-like event with optical polarimetry. The NOT $BVRI$ imaging polarimetry is low level ($\lesssim0.5$\%), and consistent with zero polarisation $+6$ -- $14$\,d post-discovery.  The event does not show the early, high, wavelength-dependent polarisation seen in AT\,2018cow \citep{Smith2018a, Maund2023}, but given our first observation is taken after this was seen in AT\,2018cow, it is plausibly not covered by our dataset.  However, both events showed values consistent with zero polarisation at $\gtrsim+8$\,d, during the phase when the optical photosphere is thought to be located in the near-relativistic outflow. With a lack of evidence for preferential orientation, it is unlikely that the two events would have been observed from similar viewing angles. As such, the fact that both of them show low polarisation implies that the high-velocity outflows in these systems are consistent with a high degree of spherical symmetry. While a number of Cow-like transients have shown evidence for highly aspherical ejecta geometry previously, it is possible that the outflow has flown around the suggested equatorial torus, thus explaining the spherical photospheres.

Obtaining high-quality optical polarimetry of more Cow-like transients is critical in investigating their ejecta geometry. These observations can be used to probe the spherical geometry of the outflow-dominated phase for the class at large, and also constrain the shape of the equatorial disk/torus. This will yield crucial (and novel) constraints on current models of Cow-like transients, making an important contribution to our understanding of the nature of these enigmatic objects.

\section*{Acknowledgements}
We thank the anonymous referee for their helpful feedback.
MP, JL and DON acknowledge support from a UK Research and Innovation Fellowship (MR/T020784/1).
TLK acknowledges support from the Turku University Foundation (grant no. 081810).
HK and TN were funded by the Research Council of Finland projects 324504, 328898, and 353019.
PC and RK acknowledge support via the Research Council of Finland (grant 340613).
BW acknowledges the UK Research and Innovation's (UKRI) Science and Technology Facilities Council (STFC) studentship grant funding, project reference ST/X508871/1. 
DC acknowledges support from the Science and Technology Facilities Council (STFC) grant number ST/X001121/1
KM acknowledges support from the Japan Society for the Promotion of Science (JSPS) KAKENHI grant JP24KK0070 and 24H01810. The
work is partly supported by the JSPS Open Partnership Bilateral Joint Research Projects between Japan and Finland (KM and HK; JPJSBP120229923).
Based on observations made with the Nordic Optical Telescope, owned in collaboration by the University of Turku and Aarhus University, and operated jointly by Aarhus University, the University of Turku and the University of Oslo, representing Denmark, Finland and Norway, the University of Iceland and Stockholm University at the Observatorio del Roque de los Muchachos, La Palma, Spain, of the Instituto de Astrofisica de Canarias.
The data presented here were obtained  with ALFOSC, which is provided by the Instituto de Astrofisica de Andalucia (IAA) under a joint agreement with the University of Copenhagen and NOT.
The data from the Seimei telescope were taken under the program 24B-N-CT14 within the KASTOR project. The Seimei telescope at the Okayama Observatory is jointly operated by Kyoto University and the Astronomical Observatory of Japan (NAOJ), with assistance provided by the Optical and Near-Infrared Astronomy Inter- University Cooperation Program.
This research has made use of the SVO Filter Profile Service \say{Carlos Rodrigo}, funded by MCIN/AEI/10.13039/501100011033/ through grant PID2020-112949GB-I00.
The Gravitational-wave Optical Transient Observer (GOTO) project acknowledges the support of the Monash-Warwick Alliance; University of Warwick; Monash University; University of Sheffield; University of Leicester; Armagh Observatory \& Planetarium; the National Astronomical Research Institute of Thailand (NARIT); Instituto de Astrofísica de Canarias (IAC); University of Portsmouth; University of Turku. We acknowledge support from the Science and Technology Facilities Council (STFC, grant numbers ST/T007184/1, ST/T003103/1, ST/T000406/1, ST/X001121/1 and ST/Z000165/1). 
This work has made use of data from the Asteroid Terrestrial-impact Last Alert System (ATLAS) project. The Asteroid Terrestrial-impact Last Alert System (ATLAS) project is primarily funded to search for near earth asteroids through NASA grants NN12AR55G, 80NSSC18K0284, and 80NSSC18K1575; byproducts of the NEO search include images and catalogs from the survey area. This work was partially funded by Kepler/K2 grant J1944/80NSSC19K0112 and HST GO-15889, and STFC grants ST/T000198/1 and ST/S006109/1. The ATLAS science products have been made possible through the contributions of the University of Hawaii Institute for Astronomy, the Queen’s University Belfast, the Space Telescope Science Institute, the South African Astronomical Observatory, and The Millennium Institute of Astrophysics (MAS), Chile.

The Legacy Surveys consist of three individual and complementary projects: the Dark Energy Camera Legacy Survey (DECaLS; Proposal ID \#2014B-0404; PIs: David Schlegel and Arjun Dey), the Beijing-Arizona Sky Survey (BASS; NOAO Prop. ID \#2015A-0801; PIs: Zhou Xu and Xiaohui Fan), and the Mayall z-band Legacy Survey (MzLS; Prop. ID \#2016A-0453; PI: Arjun Dey). DECaLS, BASS and MzLS together include data obtained, respectively, at the Blanco telescope, Cerro Tololo Inter-American Observatory, NSF’s NOIRLab; the Bok telescope, Steward Observatory, University of Arizona; and the Mayall telescope, Kitt Peak National Observatory, NOIRLab. Pipeline processing and analyses of the data were supported by NOIRLab and the Lawrence Berkeley National Laboratory (LBNL). The Legacy Surveys project is honored to be permitted to conduct astronomical research on Iolkam Du’ag (Kitt Peak), a mountain with particular significance to the Tohono O’odham Nation.
NOIRLab is operated by the Association of Universities for Research in Astronomy (AURA) under a cooperative agreement with the National Science Foundation. LBNL is managed by the Regents of the University of California under contract to the U.S. Department of Energy.
This project used data obtained with the Dark Energy Camera (DECam), which was constructed by the Dark Energy Survey (DES) collaboration. Funding for the DES Projects has been provided by the U.S. Department of Energy, the U.S. National Science Foundation, the Ministry of Science and Education of Spain, the Science and Technology Facilities Council of the United Kingdom, the Higher Education Funding Council for England, the National Center for Supercomputing Applications at the University of Illinois at Urbana-Champaign, the Kavli Institute of Cosmological Physics at the University of Chicago, Center for Cosmology and Astro-Particle Physics at the Ohio State University, the Mitchell Institute for Fundamental Physics and Astronomy at Texas A\&M University, Financiadora de Estudos e Projetos, Fundacao Carlos Chagas Filho de Amparo, Financiadora de Estudos e Projetos, Fundacao Carlos Chagas Filho de Amparo a Pesquisa do Estado do Rio de Janeiro, Conselho Nacional de Desenvolvimento Cientifico e Tecnologico and the Ministerio da Ciencia, Tecnologia e Inovacao, the Deutsche Forschungsgemeinschaft and the Collaborating Institutions in the Dark Energy Survey. The Collaborating Institutions are Argonne National Laboratory, the University of California at Santa Cruz, the University of Cambridge, Centro de Investigaciones Energeticas, Medioambientales y Tecnologicas-Madrid, the University of Chicago, University College London, the DES-Brazil Consortium, the University of Edinburgh, the Eidgenossische Technische Hochschule (ETH) Zurich, Fermi National Accelerator Laboratory, the University of Illinois at Urbana-Champaign, the Institut de Ciencies de l’Espai (IEEC/CSIC), the Institut de Fisica d’Altes Energies, Lawrence Berkeley National Laboratory, the Ludwig Maximilians Universitat Munchen and the associated Excellence Cluster Universe, the University of Michigan, NSF’s NOIRLab, the University of Nottingham, the Ohio State University, the University of Pennsylvania, the University of Portsmouth, SLAC National Accelerator Laboratory, Stanford University, the University of Sussex, and Texas A\&M University.
BASS is a key project of the Telescope Access Program (TAP), which has been funded by the National Astronomical Observatories of China, the Chinese Academy of Sciences (the Strategic Priority Research Program “The Emergence of Cosmological Structures” Grant \# XDB09000000), and the Special Fund for Astronomy from the Ministry of Finance. The BASS is also supported by the External Cooperation Program of Chinese Academy of Sciences (Grant \# 114A11KYSB20160057), and Chinese National Natural Science Foundation (Grant \# 12120101003, \# 11433005).
The Legacy Survey team makes use of data products from the Near-Earth Object Wide-field Infrared Survey Explorer (NEOWISE), which is a project of the Jet Propulsion Laboratory/California Institute of Technology. NEOWISE is funded by the National Aeronautics and Space Administration.
The Legacy Surveys imaging of the DESI footprint is supported by the Director, Office of Science, Office of High Energy Physics of the U.S. Department of Energy under Contract No. DE-AC02-05CH1123, by the National Energy Research Scientific Computing Center, a DOE Office of Science User Facility under the same contract; and by the U.S. National Science Foundation, Division of Astronomical Sciences under Contract No. AST-0950945 to NOAO.

%%%%%%%%%%%%%%%%%%%%%%%%%%%%%%%%%%%%%%%%%%%%%%%%%%
\section*{Data Availability}
Public datasets presented in this paper may be retrieved from their respective locations, referenced in the text. A machine readable version of Table~\ref{tab:phottab} is available in the Supplementary Material for this manuscript. All other data is available from the corresponding author upon reasonable request. Upon acceptance, our spectral datasets will be uploaded to WISeREP\footnote{\url{https://www.wiserep.org}}.



\bibliographystyle{mnras}
\bibliography{bib_new,extras} % if your bibtex file is called example.bib

%\begin{thebibliography}{99}
%\end{thebibliography}

%%%%%%%%%%%%%%%%%%%%%%%%%%%%%%%%%%%%%%%%%%%%%%%%%%

%%%%%%%%%%%%%%%%% APPENDICES %%%%%%%%%%%%%%%%%%%%%

\appendix
\section{Tables and Figures}

\begin{table*}
\def\arraystretch{1.1}%
\setlength\tabcolsep{8pt}
\centering
\fontsize{9}{11}\selectfont
\caption{Log of spectra obtained of AT\,2024wpp during our observing campaign. Resolving power and wavelength ranges for NOT are based on nominal instrument performance as tabulated in the ALFOSC documentation. The slit width for KOOLS-IFU is given as the fibre size, for completeness.}
\label{tab:spectral_log}
\begin{tabular}{rrrrrrrrrr}
\hline
\multicolumn{1}{c}{Instrument} & \multicolumn{1}{c}{Grism} & \multicolumn{1}{c}{Date} & \multicolumn{1}{c}{Phase} & \multicolumn{1}{c}{Exp. time} & Airmass & \multicolumn{1}{c}{Slit width} & \multicolumn{1}{c}{$R$} & \multicolumn{1}{c}{Range} \\
& & \multicolumn{1}{c}{(UT)} & \multicolumn{1}{c}{(d)} & \multicolumn{1}{c}{(s)} & & \multicolumn{1}{c}{(arcsec)} & & \multicolumn{1}{c}{(Å)}\\
\hline \hline
Seimei/KOOLS-IFU & VPH-blue & 2024 Oct 01 16:56:59 & 5.8 & 3$\times$600 & 1.60 & 0.84 & 500 & 4600-8000 \\
NOT/ALFOSC & 4 & 2024 Oct 02 01:38:16 & 6.1 & 1200  & 1.61 & 1.3 & 277 & 3200-9600 \\
NOT/ALFOSC & 4 & 2024 Oct 05 01:37:01 & 8.9 & 1200  & 1.57 & 1.0 & 360 & 3200-9600 \\
NOT/ALFOSC & 4 & 2024 Oct 08 01:27:28 & 11.6 & 1800 & 1.58 & 1.0 & 360 & 3200-9600 \\
NOT/ALFOSC & 4 & 2024 Oct 11 03:25:08 & 14.4 & 1800 & 1.45 & 1.0 & 360 & 3200-9600 \\
NOT/ALFOSC & 4 & 2024 Oct 19 03:24:06 & 21.8 & 3600 & 1.47 & 1.3 & 277 & 3200-9600 \\
\hline
\end{tabular}

\end{table*}

\begin{table*}
\def\arraystretch{1.1}%
\setlength\tabcolsep{8pt}
\centering
\fontsize{9}{11}\selectfont
\caption{Log of NOT/ALFOSC imaging polarimetry observations presented in this paper.}
\label{tab:impol_log}
\begin{tabular}{lrlrrr}
\hline
\multicolumn{1}{c}{Date} & \multicolumn{1}{c}{Phase} & \multicolumn{1}{c}{Filters}  & \multicolumn{1}{c}{Exp. time} & Airmass & \multicolumn{1}{c}{Lunar Ill.} \\
\multicolumn{1}{c}{(UT)} & \multicolumn{1}{c}{(d)}   &                              &  \multicolumn{1}{c}{(s)}      &         &  \multicolumn{1}{c}{(\%)}  \\
\hline \hline
2024 Oct 02 01:50:31.53 & $    6.1$ & $BVRI   $ & $   100$ & $   1.54$  &  $  0$  \\
2024 Oct 05 01:49:11.63 & $    8.9$ & $BVRI   $ & $   200$ & $   1.51$  &  $  5$  \\
2024 Oct 08 01:44:42.89 & $   11.6$ & $BV     $ & $   2\times300$ & $   1.49$  &  $ 24$  \\
2024 Oct 11 02:01:29.83 & $   14.4$ & $B      $ & $   2\times450$ & $   1.45$  &  $ 53$  \\
\hline
\end{tabular}

\end{table*}

\begin{table*}
    \def\arraystretch{1.1}%
    \setlength\tabcolsep{8pt}
    \centering
    \fontsize{9}{11}\selectfont
    \caption{The results of the $BVRI$ imaging polarimetry of AT\,2024wpp.}
    \begin{threeparttable}
    
    \begin{tabular}{r r r r r r r r r}
    \hline
    \hline
%    # UT MJD phase FWHM S/N Q U P X
        \multicolumn{1}{c}{Date} &  \multicolumn{1}{c}{MJD} & \multicolumn{1}{c}{Phase} & \multicolumn{1}{c}{FWHM} & \multicolumn{1}{c}{S/N} & \multicolumn{1}{c}{$Q$}    & \multicolumn{1}{c}{$U$}    & \multicolumn{1}{c}{$P$\tnote{\bf a}} & \multicolumn{1}{c}{$\theta$}     \\
             & \multicolumn{1}{c}{(d)}  & \multicolumn{1}{c}{(d)}   & \multicolumn{1}{c}{(pixel)} &     & \multicolumn{1}{c}{(\%)} & \multicolumn{1}{c}{(\%)} & \multicolumn{1}{c}{(\%)} & \multicolumn{1}{c}{($\degree$)}     \\
\hline
\multicolumn{9}{c}{\textbf{B-band}} \\
\hline
2024 Oct 01   & $60585.1$ &  $    6.1$ & $   7.50$ & $ 218.18$ & $   0.30 \pm    0.34$ & $   0.24 \pm    0.32$ & $   0.28 \pm    0.33$ & $   19.0 \pm    24.7$ \\
2024 Oct 04   & $60588.1$ &  $    8.9$ & $   5.20$ & $ 299.09$ & $  -0.53 \pm    0.24$ & $  -0.29 \pm    0.24$ & $   0.56 \pm    0.24$ & $  -75.6 \pm    11.2$ \\
2024 Oct 07   & $60591.1$ &  $   11.6$ & $   5.60$ & $ 208.10$ & $  -0.13 \pm    0.34$ & $   0.47 \pm    0.34$ & $   0.38 \pm    0.34$ & $   52.8 \pm    20.1$ \\
2024 Oct 10   & $60594.1$ &  $   14.4$ & $   3.90$ & $ 313.05$ & $   0.10 \pm    0.23$ & $  -0.35 \pm    0.23$ & $   0.30 \pm    0.23$ & $  -37.0 \pm    17.6$ \\
\hline
\multicolumn{9}{c}{\textbf{V-band}} \\
\hline
2024 Oct 01   & $60585.1$ &  $    6.1$ & $   7.50$ & $ 178.82$ & $   0.10 \pm    0.39$ & $   0.56 \pm    0.41$ & $   0.45 \pm    0.40$ & $   40.0 \pm    20.4$ \\
2024 Oct 04   & $60588.1$ &  $    8.9$ & $   5.10$ & $ 260.89$ & $  -0.04 \pm    0.27$ & $   0.12 \pm    0.27$ & $   0.07 \pm    0.27$ & $   54.5 \pm    59.7$ \\
2024 Oct 07   & $60591.1$ &  $   11.6$ & $   4.70$ & $ 151.38$ & $   0.14 \pm    0.48$ & $  -0.32 \pm    0.47$ & $   0.21 \pm    0.47$ & $  -33.4 \pm    38.4$ \\
\hline
\multicolumn{9}{c}{\textbf{R-band}} \\
\hline
2024 Oct 01   & $60585.1$ &  $    6.1$ & $   4.00$ & $ 218.65$ & $  -0.33 \pm    0.32$ & $  -0.36 \pm    0.34$ & $   0.39 \pm    0.33$ & $  -66.3 \pm    19.3$ \\
2024 Oct 04   & $60588.1$ &  $    8.9$ & $   5.40$ & $ 213.49$ & $  -0.29 \pm    0.33$ & $   0.10 \pm    0.33$ & $   0.20 \pm    0.33$ & $   80.6 \pm    31.3$ \\
\hline
\multicolumn{9}{c}{\textbf{I-band}} \\
\hline
2024 Oct 01   & $60585.1$ &  $    6.1$ & $   3.90$ & $ 160.40$ & $   0.54 \pm    0.44$ & $  -0.78 \pm    0.45$ & $   0.84 \pm    0.44$ & $  -27.7 \pm    13.5$ \\
2024 Oct 04   & $60588.1$ &  $    8.9$ & $   4.60$ & $ 127.84$ & $  -0.30 \pm    0.55$ & $  -0.26 \pm    0.55$ & $   0.25 \pm    0.55$ & $  -69.5 \pm    39.4$ \\
\hline
\hline
    \end{tabular}
    \begin{tablenotes}
    \item[a] Bias-corrected polarisation degree. Measurements whose error is larger than the value are $1\sigma$ consistent with zero polarisation.
    \end{tablenotes}
    \end{threeparttable}    
\label{tab:impol_results}
\end{table*}

\begin{table}
    \centering
    \begin{tabular}{r r r r r r}
        \hline
        Filter & MJD & Phase & Magnitude & Uncertainty & Source \\
        \hline \hline
        L & 60574.2 & -3.9 & >18.50 & - & GOTO \\
        o & 60574.6 & -3.5 & >19.46 & - & ATLAS \\
        o & 60576.0 & -2.2 & >19.35 & - & ATLAS \\
        L & 60577.2 & -1.2 & >19.52 & - & GOTO \\
        o & 60577.3 & -1.0 & >18.12 & - & ATLAS \\
        g & 60578.4 & +0.0 & 20.78 & 0.27 & ZTF \\
        L & 60579.1 & +0.6 & 18.33 & 0.08 & GOTO \\
        g & 60579.4 & +0.9 & 17.56 & 0.05 & ZTF \\
        \vdots & & & & & \\
        J & 60600.1 & +20.0 & 19.45 & 0.07 & NOT \\
        H & 60600.2 & +20.0 & 19.27 & 0.10 & NOT \\
        L & 60602.0 & +21.7 & >18.53 & - & GOTO \\
        o & 60602.3 & +21.9 & >19.27 & - & ATLAS \\
        \hline
    \end{tabular}
    \caption{Collated UV, optical, and NIR photometry of AT\,2024wpp. The values in this table are as observed, uncorrected for Galactic extinction. Upper limits refer to the 5$\sigma$ point-source limiting magnitude. Only the first few rows are shown here, with the full table being available online in machine-readable form.}
    \label{tab:phottab}
\end{table}

\begin{figure*}
    \centering
    \includegraphics[width=0.98\textwidth]{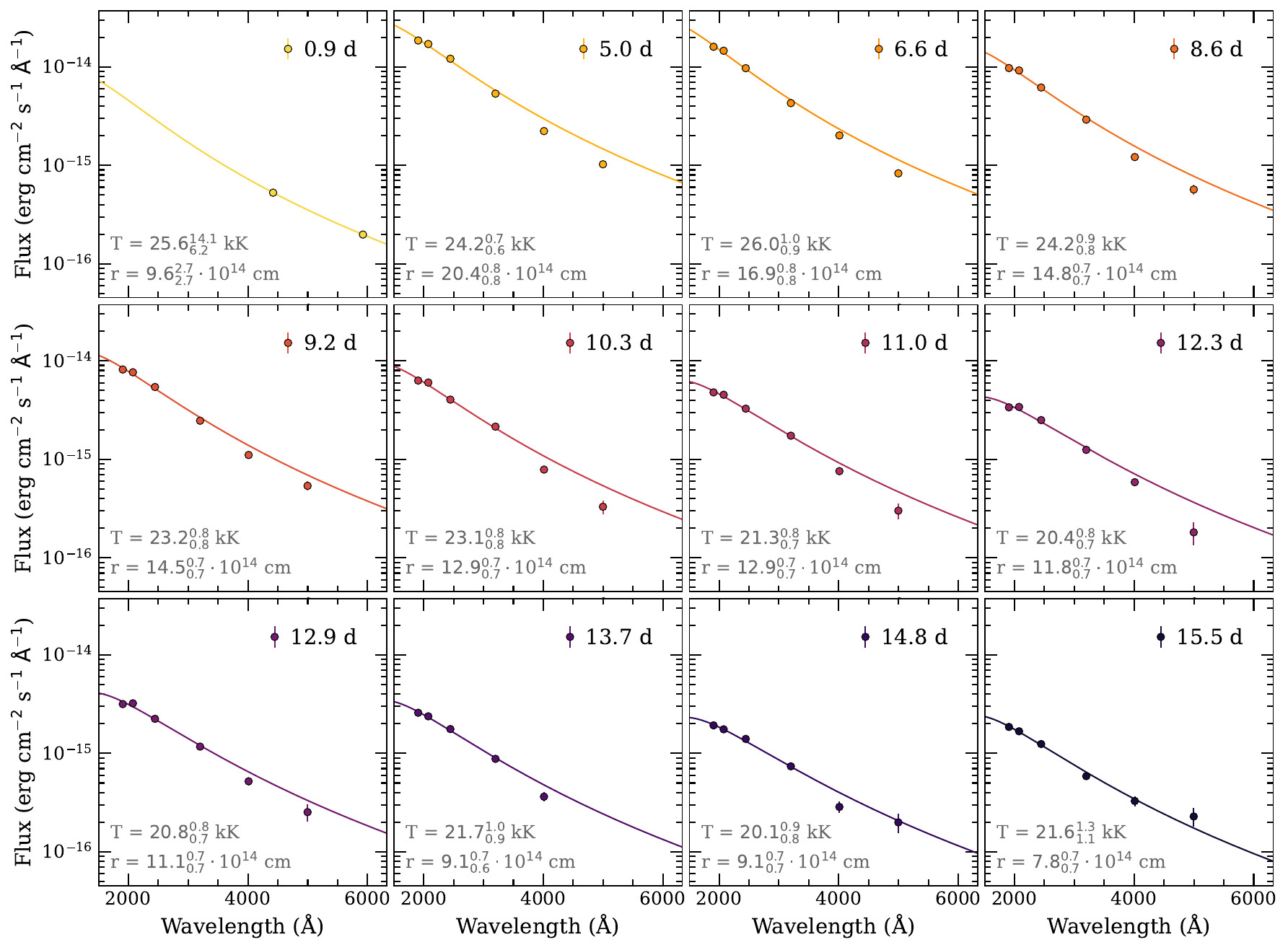}
    \caption{Blackbody fits to the multi-band photometry, including an early ZTF $gr$ epoch on the rise and six-band \textit{Swift} UVOT photometry on the decline. The shown temperature and radius errors are estimated via Monte Carlo sampling, with 10000 realisations of the data. The data is largely consistent with a blackbody of $T\gtrsim20000$\,K throughout, but the fits to \textit{Swift} photometry appear to show a slight blue excess, especially at early times. }
    \label{fig:BB_fits}
\end{figure*}

%%%%%%%%%%%%%%%%%%%%%%%%%%%%%%%%%%%%%%%%%%%%%%%%%%

% Don't change these lines
\bsp	% typesetting comment
\label{lastpage}
\end{document}